\documentclass[
 reprint,
 amsmath,amssymb,
 aps,
]{revtex4-1}
\usepackage{graphicx}
\usepackage{dcolumn}
\usepackage{bm}
\usepackage{textgreek}

\begin{document}

\preprint{APS/123-QED}

\title{
Phase field model for viscous inclusions in anisotropic networks
}

\author{Aakanksha Gubbala}
\email{A.G. and A.M.J. contributed equally to this work}
\affiliation{
Department of Chemical Engineering, Stanford University, Stanford, CA 94305\relax
}
\author{Anika M. Jena$^*$}
\affiliation{
Department of Chemical Engineering, University of California, Santa Barbara, Santa Barbara, CA 93106
}
\author{Daniel P. Arnold}
\affiliation{
Department of Chemical Engineering, University of California, Santa Barbara, Santa Barbara, CA 93106
}
\author{Sho C. Takatori}
\email{stakatori@stanford.edu}
\affiliation{
Department of Chemical Engineering, Stanford University, Stanford, CA 94305
}

\begin{abstract}
    The growth of viscous two-dimensional lipid domains in contact with a viscoelastic actin network was recently shown to exhibit unusual lipid domain ripening due to the geometry and anisotropy of the actin network [Arnold \& Takatori. Langmuir. \textbf{40}, 26570-26578 (2024)]. 
    In this work, we interpret previous experimental results on lipid membrane-actin composites with a theoretical model that combines the Cahn-Hilliard and Landau-de Gennes liquid crystal theory.
    In our model, we incorporate fiber-like characteristics of actin filaments and bundles through a nematic order parameter, and elastic anisotropy through cubic nematic gradients. 
    Numerical simulations qualitatively agree with experimental observations, by reproducing the competition between the thermodynamic forces that coarsen lipid domains versus the elastic forces generated by the surrounding actin network that resist domain coarsening.
    We observe a decrease in the growth of domain sizes, finding $R(t) \sim t^{\alpha}$ with $\alpha < 1/4$ for different actin network stiffnesses, in sharp contrast to the $\sim t^{1/3}$ scaling for diffusive growth of domains in the absence of the actin network.
    Our findings may serve as a foundation for future developments in modeling elastic ripening in complex systems.
\end{abstract}

\maketitle

\section{Introduction}
The study of viscous droplets in viscoelastic environments has drawn significant attention because of its relevance in biological systems like mammalian cells, where biomolecular condensates nucleate and coarsen within a cytoskeleton network. \cite{kasza_cell_2007, phillips_physical_2012, brangwynne_phase_2013}
More recently, the nucleation and coarsening of lipid domains along two-dimensional surfaces have been studied, inspired in part by lipid, protein, and macromolecular domains along the mammalian cell membrane that interplay with the underlying actin cytoskeleton. \cite{stanich_coarsening_2013, honigmann_lipid_2014, vogel_control_2017, yuan_membrane_2021, arnold_active_2023, arnold_lipid_2024}
The viscous lipid domains are embedded in a heterogeneous network of actin bio-polymers that exert elastic forces on the domains, modifying their structure and growth. \cite{arnold_lipid_2024, nossal_elasticity_1988, janmey_mechanical_1994}
Domain anisotropy inherently arises from interactions with networks of semiflexible actin filaments. \cite{mackintosh_actin_1997, tharmann_viscoelasticity_2007}

To investigate actin-membrane interactions, we have previously considered a 2D reconstituted system of a cushioned supported lipid bilayer as a simplified model of the plasma membrane.
At room temperature, multi-component membranes can phase separate into liquid-ordered (Lo) domains dispersed in a liquid-disordered (Ld) phase. \cite{arnold_bio-enabled_2023, vogel_control_2017, veatch_separation_2003} 
Ld-favoring lipids with positively-charged head groups may be added to the membrane to preferentially adsorb negatively-charged actin filaments to the Ld phase. \cite{arnold_bio-enabled_2023, schroer_charge-dependent_2020, honigmann_lipid_2014, vogel_control_2017}
In the presence of myosin molecular motors and ATP in the bulk fluid, the motor-driven motion of the actin filaments can accelerate the coarsening of membrane domains \cite{arnold_active_2023} and create significant spatiotemporal reorganization of lipids. \cite{gubbala_dynamic_2024}

In the absence of activity and strong hydrodynamic effects, the competition between the thermodynamic driving forces of phase separation and the elastic forces exerted by the actin network can produce unusual lipid shapes and coarsening kinetics. 
For example, unlike round, circular lipid domains observed in the absence of actin, we have previously observed lipid domains that adopt angular features with sharp cusps, such as triangles and tactoids.\cite{arnold_bio-enabled_2023, arnold_lipid_2024} 
These structure are also distinct from the bi-continuous structures observed in polymer viscoelastic mixtures undergoing spinodal decomposition. 
Furthermore, the elastic forces generated by the actin filaments hinder lipid diffusion between the domains and the bulk phase, suppressing domain coarsening. \cite{arnold_lipid_2024} 

Prior theoretical approaches in this area have considered the growth of viscous droplets in linear elastic materials \cite{larche_linear_1973, onuki_long-range_1989, onuki_phase_2001, onuki_eshelbys_1991, style_surface_2014}, as well as in polymers and gels. \cite{style_liquid-liquid_2018, rosowski_elastic_2020, curk_phase_2023, meng_heterogeneous_2024}
Recent studies show that the viscoelastic
phase can reorganize to form space-spanning networks that exert heterogeneous forces on the droplets.\cite{tanaka_viscoelastic_1997, onuki_phase_1988, onuki_theory_1989, taniguchi_network_1996, mackintosh_actin_1997, doi_gel_2009}
Consequently, droplet growth in these systems slows down, with theoretical growth exponents much lower than the $1/3$ exponent expected for Ostwald ripening and domain coalescence. \cite{meng_heterogeneous_2024, curk_phase_2023, russel_colloidal_1989}
However, these phase separation theories treat the viscoelastic
phase as isotropic, and neglect the directional microstructure of the bio-polymer.

\begin{figure}[!t]
    \includegraphics[width=\linewidth]{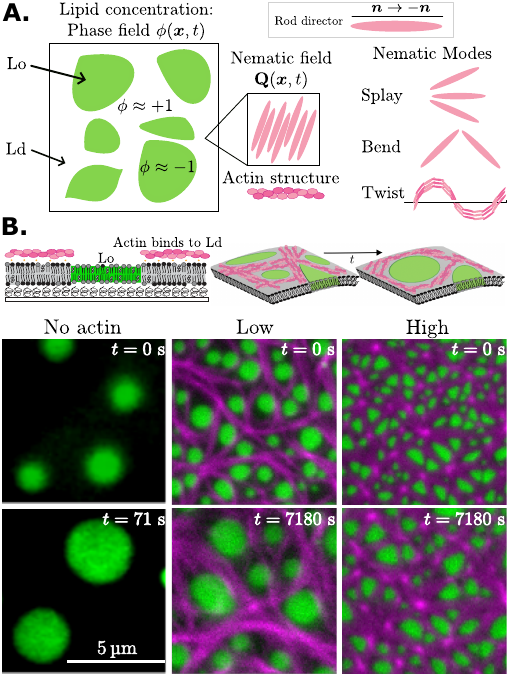}
    \caption{
    Phase-field model for a two-dimensional surface of lipid domains embedded within an anisotropic actin network.
    (A)
    The phase separation of multicomponent lipid membranes are modeled by a scalar field, $\phi(\boldsymbol{x},t)$, representing the concentration of viscous droplets (liquid ordered domains; ``Lo'') with $\phi < 0$ and a background phase (liquid disordered; ``Ld'') with $\phi > 0$. 
    The background Ld phase contains a network of actin filaments and bundles that is represented by a tensor nematic order parameter, $\mathbf{Q}(\boldsymbol{x},t)$, which allows the actin network to splay, bend, and twist out of plane. 
    (B) 
    \emph{Top:} Reconstituted lipid membrane phase separates into Lo domains (green) dispersed in a continuous Ld (black) phase on a cushioned substrate. 
    Actin filaments (pink) are electrostatically adsorbed to the Ld phase, causing the Lo domains to nucleate and coarsen against a background viscoelastic network.
    \emph{Bottom:} Experimental snapshots for Lo domains (green) in Ld phase (black) for varying densities of actin (magenta), with the top row showing early-time domains at $t = 0$ s and the bottom row showing domains at late times.
    In absence of actin, membrane domains grow to macroscopic sizes over $\sim\mathcal{O}$(min) timescales (71s in the snapshot) through Ostwald ripening and coalescence. 
    In the presence of an actin networks, domains can develop triangular shapes and sharp corners as actin elasticity resists the forces of domain coarsening.
    The resulting domains grow slowly over $\sim\mathcal{O}$(hrs) timescales.
    Scale bar is $5$ \textmu m.
    }
    \label{fig:exp-snapshots}
\end{figure}

To account for the anisotropy of filamentous actin networks, we develop a theoretical model in which the actin network is treated as a liquid crystalline material (Fig.~\ref{fig:exp-snapshots}A). 
Our model combines the Cahn-Hilliard phase separation dynamics \cite{cahn_free_1958, cahn_spinodal_1961} with an evolution equation for the nematic order parameter. \cite{gennes_physics_1993}
Previous studies have shown that actin-myosin systems exhibit liquid crystalline behavior such as topological defects, with the theory of active nematics successfully predicting the flow and deformation of such systems. \cite{caballero_vorticity_2023, colen_machine_2021, adar_permeation_2021, zhang_interplay_2018}
Moreover, dynamically evolving foams and networks, where network bonds continuously merge and break up, have emerged from models of active nematics coupled to density fields. \cite{mirza_theory_2024, maryshev_dry_2019, maryshev_pattern_2020}

Although liquid crystal theory predicts various morphologies, such as tactoids \cite{weirich_liquid_2017, schimming_equilibrium_2022}, stripes \cite{caballero_vorticity_2023}, and Schlieren textures \cite{gennes_physics_1993}, it remains unclear whether a stable network structure, such as that observed in actin, can be formed.
Another interesting aspect is the dimensionality.
Mathematically, the confined two-dimensionality imposes constraints on nematic deformation, producing fundamentally different elastic properties than are observed in 3D systems \cite{beris_thermodynamics_1994}.
In this work, we show that our 2D nematic theory serves as a minimal, yet effective model for membrane-actin interactions.
Our numerical results reveal that the nematic elasticity qualitatively models the stresses exerted by actin networks.
Additionally, the elasticity of the network can be modulated by actin density.
The actin density-dependent elastic stresses are sufficiently strong to hinder the coarsening of membrane domains, with growth exponents $\alpha < 0.25$ for varying densities.

\section{Methods}
\subsection{Theory}
We aim to develop a phase field theory that models the coupling between anisotropic elastic forces and lipid membrane phase separation.
Here, we define the scalar order field $\phi(\boldsymbol{x},t)$ as the lipid concentration, where $\phi < 0$ corresponds to the Lo phase and $\phi > 0$ corresponds to the Ld phase. 
The total free energy of the membrane-actin system in dimensionless form is given below:
\begin{equation}\label{eq:free_energy_total}
    F = F_\phi + \int \left( f_{\mathrm{bulk}} + f_{\mathrm{elastic}} + f_{\mathrm{anch}}  \right) \ d\boldsymbol{x} \ . 
\end{equation}
The phase separation of the Lo-Ld mixture is described by the Cahn-Hilliard free energy \cite{cahn_free_1958,cahn_spinodal_1961}:
\begin{equation}\label{eq:free_energy_CH}
    F_\phi = \int \left( \frac{\phi^4}{4} - \frac{\phi^2}{2} + \frac{\lambda}{2}|\nabla\phi|^2 \right) \, d\boldsymbol{x}\ , 
\end{equation}
where the first two terms represent a double-well potential, and the square gradient term models the line tension of the Lo-Ld interface.

The Ld phase is coupled to actin with nematic alignment and elasticity, distinguishing it from the actin-free viscous, isotropic Lo phase (Fig.~\ref{fig:exp-snapshots}A). 
We model actin as a bio-polymer with liquid crystalline properties using a nematic order parameter. \cite{doostmohammadi_active_2018, zhang_interplay_2018}
In two dimensions, the nematic order parameter $\mathbf{Q}(\boldsymbol{x},t)$ is defined as
\begin{equation}\label{eq:q_def}
    Q_{ij} = S\left(n_i n_j - \frac{1}{2}\delta_{ij} \right) \ , 
\end{equation}
where $S$ is the saturation value of $Q_{ij}$ in the nematic phase and $\boldsymbol{n}$ is the director field \cite{gennes_physics_1993}.
We use Einstein notation for tensor indices, in which repeated indices imply summation over that index.
Due to the symmetry and traceless properties of the $\mathbf{Q}$ tensor, we are left with only two independent fields in 2D: $Q_{xx}$ and $Q_{xy}$. 

The bulk free energy, which controls the isotropic-to-nematic phase transition, is derived from the Landau-de Gennes theory \cite{beris_thermodynamics_1994}, given by
\begin{equation}\label{eq:f_bulk_q}
    f_{\mathrm{bulk}} = \frac{\beta}{4}(Q_{ij}Q_{jk}Q_{kl}Q_{li}) - \frac{1 + \phi}{2}(Q_{ij}Q_{ij})\ , 
\end{equation}
where $\beta$ determines the nematic phase saturation. 
Using a constant saturation value for the tensor components i.e. $Q_{ij} \sim S$, Eq.~\ref{eq:f_bulk_q} models a double-well potential.
In regions where $\phi = 1$ (i.e., actin-rich Ld phase), the critical roots of Eq.~\ref{eq:f_bulk_q} have a finite value of $S = 1/\sqrt{\beta}$, indicating orientational order. 
In contrast, where $\phi = -1$ (i.e., no-actin Lo phase), the nematic field becomes zero, indicating an isotropic phase.  

The elasticity in the nematic phase arises from the gradient energy of $\mathbf{Q}$, which introduces penalties for deformation of the nematic field. 
This deformation penalty, in turn, influences the concentration $\phi$ by imposing an energetic constraint. 
This feedback loop captures the interplay between the thermodynamic driving force for phase separation and the elastic forces exerted by the actin network. 
The elastic energy is given by
\begin{equation}\label{eq:f_elastic}
    f_{\mathrm{elastic}} = \frac{1}{2} E_1 \, \frac{\partial Q_{jk}}{\partial x_i} \frac{\partial Q_{jk}}{\partial x_i} +  \frac{1}{2} E_3 \, Q_{ij}\frac{\partial Q_{kl}}{\partial x_i} \frac{\partial Q_{kl}}{\partial x_j} \ , 
\end{equation}
where we number the dimensionless elastic constants according to convention \cite{beris_thermodynamics_1994}.
Other quadratic energy terms that also respect rotational invariance are $(\partial Q_{jk}/\partial x_i)(\partial Q_{ik}/\partial x_j)$ and $(\partial Q_{ik}/\partial x_i)(\partial Q_{jk}/\partial x_j)$.
In 2D, their functional derivative is equivalent to that of the $E_1$ term, so the two terms are redundant and we retain only the $E_1$ term in Eq.~5 (see Supplementary Information).
The differences among the quadratic terms becomes apparent in 3D where they include penalty for twist deformations.
The elastic free energy terms provide control on the possible modes of nematic deformation: splay ($k_{11}$), twist ($k_{22}$), and bend ($k_{33}$), where $k_{ii}$ are the non-negative elastic constants in the director field representation. 
The twist mode is irrelevant in 2D systems, as it manifests as a helical out-of-plane deformation. \cite{wright_crystalline_1989, shendruk_twist-induced_2018}
To modify the contributions of the bend and splay modes, we first set bounds on $E_1$ and $E_3$. 
We relate $k_{ii}$ and $E_i$ through the following expressions \cite{beris_thermodynamics_1994}:
\begin{equation}\label{eq:k_E_relations}
    k_{11} = 2 E_1 S^2 - \frac{2}{3} E_3 S^3 , \quad k_{33} = 2 E_1 S^2 + \frac{4}{3} E_3 S^3 \ .
\end{equation}
Given the constraints $k_{ii}\geq 0$ and $E_1 \geq 0$, the bounds on $E_3$ are $-1.5 S \leq E_3/E_1 \leq 3/S $. 
The lower bound corresponds to a pure splay ($k_{11}$), while the upper bound corresponds to a pure bend ($k_{33}$).

Lastly, the anchoring energy is derived by considering the alignment of the director field with the interface normal, $\nabla\phi$. 
In the tensor representation of the nematic, it is given by \cite{cates_theories_2018, araki_nematohydrodynamic_2004, sulaiman_lattice_2006, chi_surface_2020}
\begin{equation}\label{eq:f_anch}
    f_{\mathrm{anch}} = \frac{\chi}{2S} \, Q_{ij} \frac{\partial \phi}{\partial x_i} \frac{\partial \phi}{\partial x_j} \ , 
\end{equation}
where $\chi$ is the dimensionless anchoring strength. 
In our experiments, we observed parallel configurations of the actin filaments conforming to the shape of the Lo phase, so we set $\chi > 0$ to enforce parallel anchoring in our model.

The evolution equations are derived by enforcing mass conservation through a conserved equation for the scalar field $\phi(\boldsymbol{x},t)$ (corresponding to Model B in the Hohenberg-Halperin classification \cite{hohenberg_theory_1977}) and by considering non-conserved dynamics for the tensor field $\mathbf{Q}(\boldsymbol{x},t)$ (Model A).
The resulting equations (Model C) are presented below:
\begin{equation}\label{eq:evolution_eq_def}
    \frac{\partial \phi}{\partial t} = \gamma \, \nabla^2 \frac{\delta F}{\delta \phi}, \quad \frac{\partial \mathbf{Q}}{\partial t} = -\frac{\delta F}{\delta \mathbf{Q}} \ .
\end{equation}
Here, $\gamma = t_Q / t_\phi$ is the ratio of two characteristic timescales in the system (see Supplemental Information for non-dimensionalization).
The nematic timescale $t_Q$ is a rotational ``viscosity" which controls the resistance to nematic orientations, and the concentration timescale $t_\phi$ is the time taken to diffuse across the Lo-Ld interface.
The expanded forms of these coupled equations can be expressed as:
\begin{align}
    \frac{\partial \phi}{\partial t} &= \gamma \nabla^2 \left[ f_{\mathrm{eff}}^\prime - \frac{\partial}{\partial x_j}\left(D^{\mathrm{eff}}_{ij}\frac{\partial \phi}{\partial x_i} \right) \right] , \label{eq:phi_eq} \\
    \frac{\partial Q_{ij}}{\partial t} &= [ 1 + \phi - \beta \, Q_{kl}Q_{kl}] Q_{ij} - \frac{\chi}{2S} \left( \frac{\partial \phi}{\partial x_i} \frac{\partial \phi}{\partial x_j} - \frac{1}{2}\delta_{ij} |\nabla \phi|^2 \right) \nonumber \\
    & + E_1 \, \nabla^2 Q_{ij} + \frac{1}{2}E_3 \Bigg( \frac{\partial Q_{kl}}{\partial x_i} \, \frac{\partial Q_{kl}}{\partial x_j} - 2 \, \frac{\partial Q_{ij}}{\partial x_k} \, \frac{\partial Q_{kl}}{\partial x_l}  \nonumber \\
    & \quad - 2 \, Q_{kl} \, \frac{\partial^2 Q_{ij}}{\partial x_k\partial x_l} - \frac{1}{2}\delta_{ij} \frac{\partial Q_{ml}}{\partial x_k}\frac{\partial Q_{ml}}{\partial x_k} \Bigg) \ , \label{eq:q_eq}
\end{align}
where the gradient operator in index notation is $\nabla \equiv \partial / \partial x_i$.
Equations \ref{eq:phi_eq} and \ref{eq:q_eq} are our main theoretical result. 
Interestingly, Eq.~\ref{eq:phi_eq} takes the form of a modified Cahn-Hilliard equation. 
Here, the usual isotropic surface tension $\lambda$ is replaced by an anisotropic tensor $D^{\mathrm{eff}}_{ij} = (\chi/S) Q_{ij} + \lambda \delta_{ij}$ and the bulk free energy is modified to $f_{\mathrm{eff}}^\prime = \phi^3 - \phi - Q_{ij}Q_{ij} / 2$.
The isotropic line tension parameter is defined as $\lambda = 1 + |\chi| / 2$, where $\chi$ is the residual contribution from transforming a director field description of anchoring into a nematic field framework (see Supplemental Information for details).

To provide a physical understanding of the model, we focus on each term in Eq.~\ref{eq:q_eq}. 
The $\chi$ term in Eq.~\ref{eq:q_eq} represents the interfacial stress derived from the gradient energy of $\phi$ \cite{cates_theories_2018}. 
We can interpret $\mathbf{Q}$ as a strain tensor, where its time-evolution corresponds to a rate-of-strain tensor, which is equal to the sum of higher-order variants of strain and an interfacial stress. 
Thus, the evolution equation of $\mathbf{Q}$ is analogous to a continuum mechanics description of a viscoelastic material. 
Through the liquid crystal framework, we have effectively coupled the elastic and concentration fields within an Eulerian frame of reference.

We perform linear stability analysis on a horizontally oriented nematic phase to gain more mechanistic insight (see Supplemental Information).
The anisotropic line tension $\boldsymbol{D}^{\mathrm{eff}}$ introduces hyperbolic gradients of the form $k^2(k_x^2 - k_y^2)$ in Eq.~\ref{eq:phi_eq}, causing the longitudinal and transverse $\phi$ field perturbations to grow at different rates. This is in contrast to the elliptic nature of the $\nabla^4$ (or $k^4$) term in regular Cahn-Hilliard equations, where the perturbations are dampened radially.
Standard coarsening models such as the Lifshitz-Slyozov theory neglect the contribution of the fourth order surface tension $\nabla^4\phi$, as it relaxes much faster than the diffusive $\nabla^2 f^\prime_{\mathrm{eff}}$. \cite{bray_lifshitz-slyozov_1995}
By assuming the Gibbs-Thomson boundary condition, the second order gradients are sufficient to predict the $\sim t^{1/3}$ growth rate.
In our model, we cannot neglect line tension and must investigate the complete non-linear dynamics of Eqs.~\ref{eq:phi_eq}-\ref{eq:q_eq} to understand the coarsening behavior.

We numerically evolve the fields $\phi(\boldsymbol{x},t)$, $Q_{xx}(\boldsymbol{x},t)$, and $Q_{xy}(\boldsymbol{x},t)$ from an initially homogeneous state with added noise by solving Eqs.~\ref{eq:phi_eq} and \ref{eq:q_eq} using Fourier pseudospectral methods on a $512^2$ grid with periodic boundary conditions (see Supplemental Information for numerical implementation). 
To reduce the number of parameters, we set $(\beta, \chi, \gamma) = (1, 1, 0.1)$ and $E_3 = E_1$, which ensures numerical stability (see Supplemental Information for linear stability analysis).
We are interested in modeling the semi-flexible nature of actin filaments, which can be achieved by modulating the elastic constants $E_1$ and $E_3$.
However, it is important to note that the ratio $E_3/E_1$ is confined to a narrow range (see Eq.~\ref{eq:k_E_relations}).
Hence, we choose $E_3/E_1 = 1$ such that the ratio of the bend and splay constants is $k_{33}/k_{11} = 2.5$; we impose a larger energetic penalty on the bending of the nematic material compared to splaying.
We vary $E_1$ and $\phi_o$ parameters to modulate the stiffness of the nematic actin field, where $\phi_o = \frac{1}{V}\int_{V} \phi(\boldsymbol{x}, t) d\boldsymbol{x}$ is the average value of $\phi$ calculated over a region of volume $V$, and determines the overall composition of the phases. 

We compare our theoretical model to a simplified experimental realization of a membrane-actin material.
In simulations, we assume low actin density leads to a lower elastic energy so we choose $(E_1, \phi_o) = (0.1, 0.3)$.
For high actin density samples, we choose $(E_1, \phi_o) = (2, 0.75)$; there is a larger elastic penalty, coupled with a high concentration of the nematic phase, which effectively reduces mass transport between the domains and Ld phase. 
Like most phase field theory efforts, our goal is to capture qualitative features of the experiments with the aim of predicting physical mechanisms, as opposed to precisely mapping quantitative parameters.

\begin{figure}[!t]
    \includegraphics[width=\linewidth]{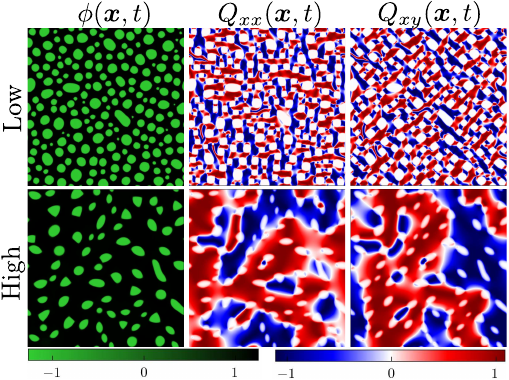}
    \caption{
    Numerical solutions to Eqs.~\ref{eq:phi_eq} and \ref{eq:q_eq} show that nematic elasticity modulates the structure and growth of 2D viscous domains.
    Domain phase field $\phi$ (\emph{left}) phase separates into viscous droplets (green) dispersed in a nematic background (black), for low (\emph{top}) and high actin density (\emph{bottom}).
    Positive (red) and negative (blue) regions of the nematic tensor components $Q_{xx}$ (\emph{center}), $Q_{xy}$ (\emph{right}) fields are interwoven like networks.
    Increasing the strength of nematic elasticity creates sharp cusps in viscous domains, which qualitatively agrees with the morphologies observed in our experiments (see Fig.~\ref{fig:exp-snapshots}B).
    White regions correspond to isotropic phase, where $Q_{ij} \approx 0$.
    Model parameters used for the simulation are $(E_1, \phi_o)=(0.1, 0.3)$ for low actin density and $(E_1, \phi_o)=(2, 0.75)$ for high actin density.
    }
    \label{fig:simulation}
\end{figure}
    
\subsection{Experiments}
Giant unilamellar vesicles (GUVs) are prepared from a lipid solution composed of 45.7\% dioleoylphosphatidylcholine (DOPC), 35\% dipalmitoylphosphatidylcholine (DPPC), 15\% cholesterol, 4\% positively-charged dioleoyl-3-trimethylammonium propane (DOTAP), and 0.3\% 1,2-distearoyl-sn-glycero-3-phosphoethanolamine-N-[poly(ethylene glycol)2000-N'-carboxyfluorescein] (DSPE-PEG2k-FITC) using electroformation. \cite{angelova_liposome_1986}
This mixture of lipids and cholesterol phase-separates into liquid-ordered (Lo) domains dispersed within a liquid-disordered (Ld) continuous phase.
The DOPC and DOTAP lipids enrich into the liquid-disordered (Ld) continuous phase, while DPPC and fluorescently-labeled DSPE-PEG2k-FITC enrich into liquid-ordered (Lo) domains.

The GUVs are deposited on a low-friction polymer-cushioned coverslip, where they rupture into planar bilayers due to electrostatic attractions between the positively charged DOTAP and negatively charged glass substrate as described previously \cite{arnold_lipid_2024}.
Coverslips are heated to 37\textdegree C for 15 minutes to melt Lo domains.
Filamentous actin (F-actin) is electrostatically adsorbed to the positively-charged DOTAP lipids in the Ld phase (see Supplemental Information for actin preparation).
When the membrane is quenched to room temperature, the Lo and Ld phases separate once again and the actin partitions into the DOTAP-enriched Ld phase.
Meanwhile, the Lo domains nucleate and coarsen within the pores created by the actin network.
The actin filaments bind to the membrane, but do not incorporate into the membrane.
Thus, lipids can still exchange between domains by dissolving into and diffusing through the Ld phase before re-condensing into a different Lo droplet.
However, entire Lo domains are generally unable to cross the regions of Ld phase pinned to the actin filaments (Fig.~\ref{fig:exp-snapshots}B).

Despite our best efforts to eliminate substrate friction using a cushioned polymer layer, the experimental protocol introduces some frictional contacts between the membrane and substrate.
This creates a large variability in domain growth rate.
To account for the heterogeneous distribution of actin on the membrane, we divided our samples into two groups: low and high actin density. 
We calculate a relative actin density by dividing the mean fluorescence intensity of actin by the available area of the Ld phase\cite{arnold_lipid_2024} (Supplemental Fig.~S1).
To classify our samples, we determine a cut-off for the actin density by comparing the domain structures of each sample.
We observe that at low densities, actin aggregates to form thin bundles around Lo domains.
The domains remain nearly circular as the actin network exerts minimal forces at the domain interface (Fig.~\ref{fig:exp-snapshots}B).
At high actin densities, the filaments completely occupy the Ld phase.
The rigidity of the actin filaments dominates the line tension of the Lo-Ld interface, forcing the domains to conform to the shape of the network (Fig.~\ref{fig:exp-snapshots}B).

\begin{figure}[!t]
    \includegraphics[width=\linewidth]{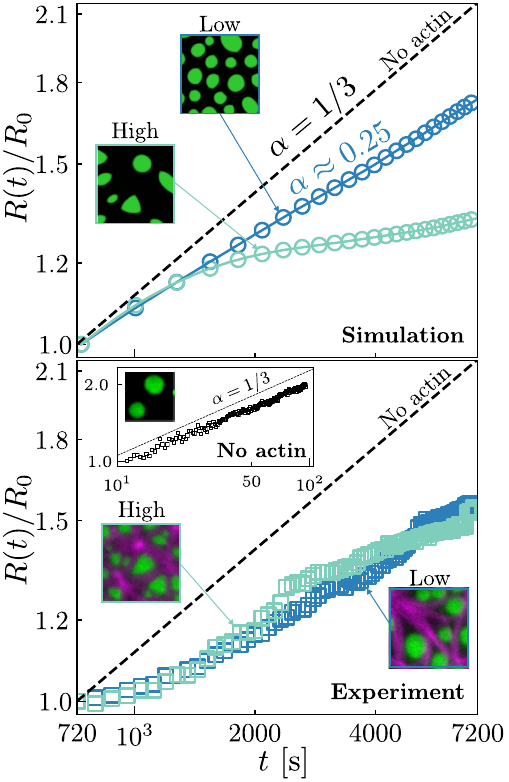}
    \caption{Average size of lipid domains, $R(t)$, is plotted as a function of time $t$ for experiments (squares, bottom) and numerical solutions of Eqs.~\ref{eq:phi_eq} and \ref{eq:q_eq} (circles, top).
    In absence of actin, domain coarsening follows the classical $\alpha=1/3$ growth (black dotted line, bottom inset). \cite{arnold_active_2023, veatch_separation_2003}
    In contrast, domain coarsening is hindered to $\alpha \approx 0.25$ in the presence of low actin density (dark blue) and to $\alpha < 0.25$ with high actin density (light blue).
    We fit a power law scaling, $R(t) \sim t^\alpha$, to highlight the log-linear coarsening trends in the low actin density simulations (dark blue circles).
    In Eqs.~\ref{eq:phi_eq} and \ref{eq:q_eq}, model parameters for low actin density are $(E_1, \phi_o)=(0.1, 0.3)$ and for high actin density are $(E_1, \phi_o)=(2, 0.75)$.
    Each data set is normalized by the domain size $R_0$ measured at $720$s to capture the onset of domains filling all void spaces of the actin network and competing with the elastic forces.
    Experimental data are averaged over 3 low actin density samples and 2 high actin density samples.
    Simulation data are averaged over 4 realizations.
    }
    \label{fig:domain-growth}
\end{figure}

Comparing the present system to a control experiment without actin, we observe that the Lo domains remain circular and coarsen as $\sim t^{1/3}$, which is expected for Ostwald ripening and coalescence (Fig.~\ref{fig:exp-snapshots}B).\cite{arnold_active_2023}
Following a quench, small domains nucleate within seconds and grow to $\approx 1-10$ \textmu m over the course of a few minutes (see Fig.~\ref{fig:domain-growth} inset).
In contrast, in the presence of the actin network, small domains nucleate but coarsen very slowly over $\approx 2$ hours (Fig.~\ref{fig:exp-snapshots}B).
Despite the well-established elasticity of actin \cite{phillips_physical_2012}, here, actin filaments are not cross-linked and are adsorbed to a quasi-2D viscous fluid.
Thus, as this multiphase fluid coarsens, the actin filaments rearrange and relax to accommodate coarsening.
In our previous publication\cite{arnold_lipid_2024}, we tracked the motion of fluorescent-labeled actin bundles, demonstrating that the actin is mobile within the Ld phase.
This rearrangement of individual elastic filaments causes the Ld-actin phase, as a whole, to exhibit viscoelastic character.

During the early stages of nucleation and coarsening, the domains grow rapidly to fill the empty voids within the actin network.
This decreases the available Ld area for actin adhesion, which in turn constrains the mass transfer of lipids through a frustrated actin network (Fig.~\ref{fig:exp-snapshots}B).
Eventually, the domains grow large enough to push against the surrounding actin network, which generates elastic restoring forces to hinder domain coarsening. 
Over these intermediate time scales ($\approx$ 10 min - 2 hours), there is a competition between the thermodynamic forces that drive domain expansion versus the elastic forces of the actin network that compress the domains.
The goal of our work is to develop a theory to predict the intermediate time scales of our system, when domain expansion and actin elasticity generate competing forces.
At very long time scales ($\gtrsim$ 2 hours), the actin may begin to unbind from the lipid membrane in part due to the large thermodynamic forces that drive domain expansion. We do not focus on the very late stage processes in this work.

\section{Results}
\begin{figure*}[!t]
    \centering
    \includegraphics[width=\linewidth]{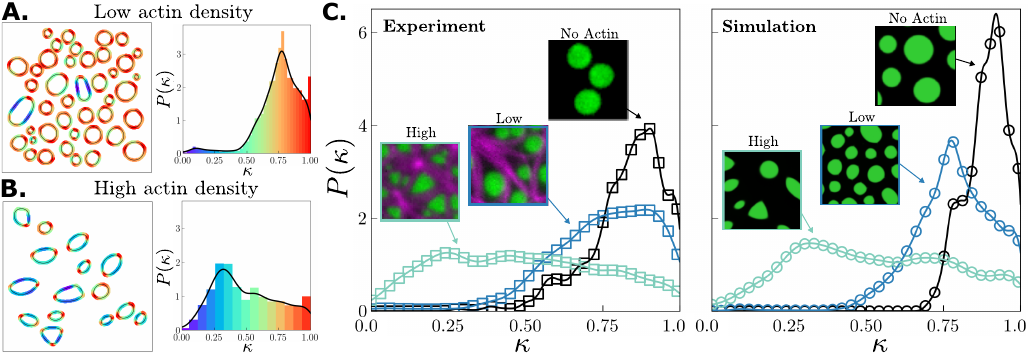}
    \caption{
    Local curvatures of lipid domains become heterogeneous with increasing actin network elasticity. 
    (A \& B) A collection of domain boundaries (\emph{left}) for the low (top) and high actin density (bottom) simulations are presented, where the domain boundary is shaded by normalized curvature, $\kappa$. 
    Histogram of $\kappa$ values is plotted (\emph{right}), where the colors are mapped to the shading of the domain boundary.
    A kernel density estimator is used to generate the smoothened probability distribution $P(\kappa)$ (solid black line).
    To emphasis curvature heterogeneity, raw curvature of individual domains is normalized by its maximum value to yield $\kappa$.
    For instance, perfectly circular domains have a constant curvature, which corresponds to a delta distribution, $P(\kappa) = \delta (\kappa = 1)$ under our definition of normalized curvature.
    As shapes deviate from perfect circles, the curvature becomes heterogeneous, with high $\kappa$ at cusps (red) and low $\kappa$ elsewhere (blue).
    (C) Probability distribution $P(\kappa)$ is plotted as a function of normalized curvature $\kappa$ for experiments (squares, \emph{left}) and simulations (circles, \emph{right}).
    Each curve corresponds to a cropped inset image highlighting representative domain structures obtained from experiments (\emph{left}) and simulated $\phi$ field (\emph{right}).
    For low actin density (dark blue), domains are slightly deformed, with peaks closer to $\kappa = 1$. 
    For high actin density (light blue), domains form triangular structures, with peaks closer to $\kappa = 0$.
    }
    \label{fig:curvature}
\end{figure*}

Solutions of Eqs.~\ref{eq:phi_eq} and \ref{eq:q_eq} are shown in Fig.~\ref{fig:simulation} for the lipid concentration field and the nematic order fields. 
In low actin density simulations, the nematic and concentration fields nucleate and coarsen over a similar time scale. 
The nematic phase forms a network around the isotropic domains, limiting diffusion between the two phases (Fig.~\ref{fig:simulation}, Supplemental Video~S2).
In high actin density simulations, the $\mathbf{Q}$ fields undergo rapid phase separation initially, creating defects that act as nucleation sites for domain growth.
The domains subsequently develop sharp triangular features as they grow within the voids generated by the defects, which are consistent with our experimental observations.
Upon reaching sufficiently large sizes, the domains begin to coarsen, and the $\mathbf{Q}$ fields become constrained by the dynamics of the $\phi$ field (Fig.~\ref{fig:simulation}, Supplemental Video~S2).

Overall, the qualitative features of the domain shape and distribution are captured by our coupled concentration-nematic phase field model.
Equations ~\ref{eq:phi_eq} and \ref{eq:q_eq} admit a solution for a stable network of nematics with embedded viscous droplets, which is distinct from prior studies that involve concentration-nematic coupling.
For example, prior studies have also shown networks of nematics driven by active forces \cite{maryshev_pattern_2020, mirza_theory_2024}.
However, there the active forces were so strong that the networks were highly transient and dynamic; thermal forces played a negligible role in the resulting patterns.
In addition, seminal prior works in phase field modeling have utilized linear elasticity theory to model solid crystalline grains embedded within a viscous solvent \cite{onuki_phase_2001, onuki_long-range_1989, nishimori_pattern_1990} and liquid crystal/gel foam-like composites \cite{uchida_phase_2001}.
However, those models are unable to predict the coarsening of viscous droplets embedded within a nematic elastic medium, which is controlled by domain diffusivity, interfacial line tension, and non-linear elastic forces exerted by the nematic phase.

We compare the performance of our model against previously measured metrics for the membrane-actin system -- the average size of domains over time and the distribution of domain curvature \cite{arnold_lipid_2024}.
Within the multiphase membrane, the nucleation and growth of domains makes the actin network softer and more viscous over time.
In Fig.~\ref{fig:domain-growth}, we focus our attention on analyzing the domain growths observed during the intermediate time scales, where the actin has sufficiently relaxed from its initial constrained configuration, allowing the domains to diffuse and grow.
To these ends, we measure the average size of domains $R(t)$, from $t=720$s to $t=7180$s.
The resulting $R(t)$ is normalized by the domain size measured at $t=720$s.
The non-dimensional simulation time is scaled to match the duration of the experiment.
For low actin density simulations, we fit the data to a power-law model, $R(t) = A t^\alpha$, finding $\alpha \approx 0.25$.
Interestingly, this result is consistent with an interfacial surface tension-driven growth  ($\alpha=1/4$) compared to a bulk diffusion-driven growth ($\alpha=1/3$) \cite{bray_lifshitz-slyozov_1995}.
In Eq.~\ref{eq:phi_eq}, we see that even a small degree of nematic ordering will induce the nonlinear surface tension, promoting an interface-driven growth.
The corresponding experimental data also exhibits a similar trend to the simulation data.
In the high actin density samples, coarsening is slower than interface-driven growth ($\alpha < 0.25$).
We avoid fitting precise exponents as the $R(t)$ data does not have a log-linear relationship with time.
This non-linearity in the high actin density case can be attributed to the slow growth of domains, which gradually softens the network and consequently the effective mobility over time. 
Overall, our results in Fig.~\ref{fig:domain-growth} confirm that increasing elastic stress can dramatically decrease the rate of domain coarsening.

Our model also qualitatively predicts the morphological features observed in experiments.
We analyze the structure of membrane domains by calculating the curvature of each domain boundary, normalized by its maximum curvature (see Supplemental Fig.~S3 for calculation details). 
This quantity, $\kappa$, is a measure of curvature heterogeneity within the system.
Circular domains have a constant curvature, resulting in $\kappa = 1$.
In contrast, irregular shapes exhibit high $\kappa$ at the corners and low $\kappa$ in the curved regions.
In Fig.~\ref{fig:curvature}, we present the probability distribution of $\kappa$ for systems with and without actin.
In the absence of actin, the domains remain nearly circular, leading to a probability distribution that is sharply peaked at $\kappa \approx 1$.
As we increase actin density, the domains become more angular and exhibit heterogeneities.
This is reflected in the curvature distribution curves, where the peak value of $P(\kappa)$ shifts towards smaller values as the number of sharp features increases.

Alternatively, one can interpret $P(\kappa)$ as a measure of stress heterogeneity within the material.
From the classical Laplace-Young equation, we know that the domain curvature is proportional to excess stress felt by the lipid domains in response to an imposed actin elasticity. \cite{style_surface_2014, arnold_active_2023}
Perfectly spherical domains experience constant stress from the background fluid, giving rise to a homogeneous stress distribution (high $P(\kappa)$), while anisotropic forces allow the stress to unevenly accumulate over the domain interface (low $P(\kappa)$).

Returning to the nematic theory, the elasticity in our model arises from the nematic gradient energy in Eq.~\ref{eq:f_elastic}.
There is a penalty for the deformation of the nematic field, which increases with the elastic constants $E_i$.
When nematic fields deviate from uniaxial ordering, singularities are generated at the sites of discontinuity, known as topological defects. \cite{blow_biphasic_2014, maryshev_pattern_2020}
These defects also correspond to regions of high $\kappa$ and consequently lead to regions of high stress accumulation due to significant elastic distortions.
Thus, there exists a mapping between stress and curvature for both experiments and simulations.
Our theoretical model naturally generates an anisotropic surface tension that captures these curvature effects and connects back to actin deformation through the nematic tensor, without the need for any ad hoc assumptions.
In Fig.~\ref{fig:curvature}C, we show that $P(\kappa)$ varies with actin density similarly for experiments and simulations, confirming that nematic elasticity captures the anisotropic deformations of the actin networks.

Our simulation results predict the suppression of coarsening but fail to generate a triangular network in the nematic phase, as shown in Fig.~\ref{fig:exp-snapshots}B. 
A network of triangles could be formed by an arrangement of $-1/2$ corner defects.
This must be compensated by positive defect charges to maintain conservation of topological charges (see Supplemental Fig.~S5 for a schematic of the defects). 
Positive defect charges generate more circular structures, making it impossible to create a network of triangles in two dimensions. 
We hypothesize that in a more realistic quasi-2D geometry, the third dimension may provide additional degrees of freedom for the nematic network to deform.
For instance, in three dimensions, the nematic can twist to form bundles and access more types of defects in the topological space.

Our experimental model of a membrane-actin composite is actually a thin three dimensional system which is $\sim\mathcal{O}(10-50\ \mathrm{nm})$ thick because the actin bundles can lie on top of each other on the membrane.
Therefore, the three-dimensionality allows the actin bundles to twist out-of-plane to dissipate in-plane stresses, unlike a purely 2D system.
This would allow an in-plane slice to form a network of triangular shapes, as the curvature from positive defects can dissipate through the helical twist contortions of actin.
However, due to the limited modes of deformation available in our 2D model, the common charge-preserving structures are tactoids \cite{schimming_equilibrium_2022, weirich_liquid_2017}, which we commonly observed in simulations (see Fig.~\ref{fig:simulation}).

\section*{Conclusions}
In this article, we develop a phase field model which combines the Cahn-Hilliard and Landau-de Gennes theories to study the dynamics of viscous droplets embedded in an elastic nematic field.
We include two non-linear effects in our model to overcome the trivial uniaxial ordering of nematics: parallel anchoring and anisotropic elasticity through cubic nematic gradients.
Through the parallel anchoring condition, we align the nematic phase along the droplet interface, which results in a modified line tension and ultimately leads to tactoid shapes.
The anisotropic elasticity introduces tunable splay and bend deformations, allowing the droplets to deviate from the stable tactoid shapes. 
Additionally, these effects impose a penalty on concentration diffusion, hindering the growth of droplets that would otherwise grow as $R(t) \sim t^{1/3}$.
Finally, we demonstrate a promising application of our model by corroborating simulations with experiments on membrane-actin composites, in which filamentary actin networks suppress the coarsening of viscous lipid domains.
Our numerical results qualitatively replicate the network morphology and the extent of coarsening suppression in experiments.

We anticipate that our framework can be extended beyond applications to the membrane-actin system described here.
For instance, by reducing the role of anisotropic elasticity, one can study the coarsening of tactoid structures, which is useful in understanding confined liquid crystalline droplets.
Tactoids are also observed in rod-like biopolymer solutions, and understanding their spatial organization may provide insights for designing novel biological liquid crystals. \cite{weirich_liquid_2017}
In another approach, we can modify the boundary conditions on the domain surface through anchoring. 
While parallel anchoring is more common, perpendicular anchoring is sometimes found in nematic colloidal suspensions, where small water droplets are dispersed in a nematic liquid crystal. \cite{poulin_novel_1997}.
By choosing $\chi < 0$, the director is allowed to orient perpendicular to the droplet surface, introducing elastic deformations that are different from the present system.
With further extension into quasi-2D and 3D geometries which permit additional modes of deformation, we expect that the phase field Model C may contribute to the understanding of complex anisotropic fluid environments.

\section*{Author contributions}
A.G. and S.C.T. conceptualized the work;
A.M.J. and D.P.A. performed experiments and analyzed experiment data;
A.G. and S.C.T. developed theory;
A.G. performed numerical simulations;
S.C.T. supervised the study;
A.G., A.M.J., D.P.A., and S.C.T. wrote the paper.

\section*{Conflicts of interest}
There are no conflicts to declare.

\section*{Data availability}
Data for this article can be generated with the simulation code available at \url{https://github.com/aakanksha-gubbala/ActinNematics}.

\section*{Acknowledgments}
We thank Austin Hopkins and Cristina Marchetti for helpful discussions on nematic field theory.
This material is based upon work supported by the National Science Foundation under Grant No.~2440029.
S.C.T. is supported by the Packard Fellowship in Science and Engineering.

%%%REFERENCES%%%
\providecommand*{\mcitethebibliography}{\thebibliography}
\csname @ifundefined\endcsname{endmcitethebibliography}
{\let\endmcitethebibliography\endthebibliography}{}

\end{document}